\documentclass[letter]{aa} 

\usepackage{graphicx}
\usepackage{txfonts}
\usepackage{verbatim}
\usepackage{natbib}
\usepackage{array}
\usepackage{threeparttable}

\def\kms{km\,s$^{-1}$}

\def\msun{M$_{\odot}$}

\def\rsun{R$_{\odot}$}

\def\l{$\lambda$}

\def\hea{He\,{\sc i}}
\def\heb{He\,{\sc ii}}

\def\fw{\textsc{fastwind} }

\def\spamms{\textsc{spamms} }

\begin{document}

   
   \title{The effects of rotation on massive star spectroscopic observables}

   \author{M. Abdul-Masih
          \inst{1}
          }

   \institute{European Southern Observatory, Alonso de Cordova 3107, Vitacura, Casilla 19001, Santiago de Chile, Chile\\
            \email{michael.abdul-masih@eso.org}
             }

   \date{}

 
  \abstract
   {Rotation is ubiquitous among massive stars.  With rotation comes a deformation to the surface geometry, which in turn leads to alterations in the distribution of parameters across the surface including surface gravity, temperature and ionization balance of surface elements.  Often, these 3D effects are neglected when analyzing spectra of rapidly rotating massive stars.}
   {We aim to determine whether neglecting the 3D deformations resulting from rapid rotation has an impact on the final spectroscopic observables, and if so to what degree.}
   {Using the \spamms code, we generate a grid of synthetic spectra that account for the 3D geometry of rapidly rotating stars and compare them to synthetic spectra generated assuming spherical geometry.  Using equivalent width and full width half maximum measurements as proxies, we determine how the measured temperature, helium abundance and projected rotation rates of individual lines in different ionization states vary with rotation rate and inclination.}
   {We find that the 3D geometry can have a significant impact on the measured parameters.  We show that the temperature is highly dependent on both the rotation rate and the inclination, and that the same system viewed at different inclinations can have measured temperatures that differ by as much as 10\%.  We also find that the helium abundance can be underestimated by as much as 60\%, and that lines in different ionization states can have measurable differences in rotation rates.  We demonstrate that these differences in rotation rates can be seen in observed data and show that this could allow for an inclination independent measurement of the rotational velocity.}
   {Our results indicate that neglecting the 3D effects of rotation can cause significant biases in the measured spectroscopic parameters, and that in many cases, the measured values are more than 3$\sigma$ away from the true values. }

   \keywords{Stars: rotation --
                Stars: massive -- 
                Stars: fundamental parameters -- 
                Stars: abundances -- 
                Techniques: spectroscopic}

   \maketitle
%

\section{Introduction}\label{sec:intro}
   
   Rotation plays a crucial role in the lives of massive stars \citep[see, e.g.,][]{Langer2012, Rivinius2013}.  One key feature of rotation is the deviation from spherical symmetry.  When a star is rapidly rotating, the centrifugal force causes the equatorial region of the star to bulge, resulting in a radius differential between the pole and the equator that can reach up to 50\% in the most extreme cases \citep{Maeder2009}.  This change in radius across the surface leads to a change in surface gravity, which in turn leads to a change in the temperature profile, which can alter the ionization balance of diagnostic elements \citep{vonZeipel1924}.  This cascading effect can have large implications on our understanding and interpretation of rapidly rotating stars.
   
   In addition to changing the surface distributions of several observable parameters, rotation has been shown to be important for internal mixing \citep[e.g., ][]{Maeder1987, Brott2011a}.  As a rotating star evolves, differential rotation between the core and the envelope leads to sheer instabilities, which can mix material between core and the near-core region of the radiative envelope \citep{Endal1978, Heger2000b}. Additionally, because a rapidly rotating star cannot be in both thermal and hydrostatic equilibrium simultaneously \citep{vonZeipel1924}, large scale circulations known as Eddington-Sweet circulations are known to operate in the envelope of these systems \citep{Eddington1925, Sweet1950}.
   
   Recent studies have demonstrated that rotation is ubiquitous in massive stars \citep[e.g., ][]{Ramirez-Agudelo2013, Bodensteiner2020b, Schootemeijer2022}, however, despite their importance, the effects of rotation are often treated in 1D for spectroscopic modeling.  Non-local thermodynamic equilibrium (NLTE) effects become important in the temperature regime that massive stars are found, and due to the complexity of solving the NLTE radiative transfer equations, most codes are computed in 1D \citep[see e.g., ][]{Puls2005, Hillier1998, Grafener2002}. For this reason, observed spectra are usually fit with 1D synthetic spectra that have been convolved with a rotation kernel.  
   
   By modeling these systems in 1D and implicitly assuming that the surface is spherical and uniform,  we not only neglect important spectroscopic features that directly result from the 3D geometry, but we may also bias our measurements \citep[see e.g., ][]{Townsend2004, Fremat2005}.  In this Letter, we demonstrate that the 3D effects of rapid rotation can have a significant impact on the stellar parameters determined via 1D spectroscopic fits and we quantify the magnitude of these effects.  Using state-of-the-art 3D spectral synthesis techniques, we generate synthetic line profiles that account for the 3D geometry of rapidly rotating stars and compare them with synthetic spectra assuming spherical geometry.  The model computations and comparison methods are described in Sect. 2.  In Sect. 3, we describe the results, and we discuss the implications of these findings in Sect. 4.  

\section{Methods}

\subsection{Computation of 3D synthetic line profiles}\label{subsec:spamms_setup}

In order to assess the how the 3D geometry of a rapidly rotating massive star affects its spectroscopic observables, we need a code that can accurately account for both the unique geometries of these systems as well as the NLTE effects that become important in this temperature regime.  For this purpose, we utilize the \spamms code \citep{Abdul-Masih2020a}. \spamms is a spectral synthesis tool that combines the advanced physics and 3D meshing techniques of the \textsc{phoebe ii} code \citep{Prsa2016, Horvat2018, Jones2020, Conroy2020} with the 1D NLTE radiative transfer code \fw \citep{Puls2005, Carneiro2016, Sundqvist2018a}.  \spamms works by first creating a 3D triangulated mesh that represents the surface geometry of a given system, and then based on the local conditions of each patch across the surface (i.e. temperature, surface gravity, radius, and radial velocity), synthetic \fw line profiles are assigned.  Finally, \spamms integrates over the visible surface resulting in the final synthetic line profiles, which account for the 3D geometry of the system as well as the distributions of the surface parameters that result from these distortions.

Using \textsc{spamms}, we generate a small grid of models at various rotation rates and inclinations, allowing only the geometry and resulting surface parameter distributions to change.  This ensures that the changes to the resulting line profiles are due solely to the 3D effects of rotation. The dimensions of our grid cover inclinations from $i = 0^{\circ}$ to $i = 90^{\circ}$ in steps of 15$^{\circ}$ and rotation rates ($v_\mathrm{rot}$) as a fraction of the linear critical or breakup velocity ($v_c$) of $\Upsilon = v_\mathrm{rot}/v_c =$ 0.001, 0.25, 0.5, 0.6, 0.7, 0.8, 0.9, 0.95 and 0.99.  All models are given a polar radius of $R_\mathrm{pole} = 6$\rsun{}, a mass of $M = 25$\msun{} and a reference temperature of $T_\mathrm{ref} = 40$kK.  Further, we set the helium abundance to be $N_\mathrm{He}/N_\mathrm{H} = 0.2$ where $N_\mathrm{He}$ and $N_\mathrm{H}$ are the number abundances of helium and hydrogen respectively.  For each model we generate two synthetic line profiles: \hea\ \l 4471 and \heb\ \l 4541, as these are typically used as temperature diagnostics in massive stars. 

Given that the change in geometry leads to a nonuniform distribution of temperatures across the surface of a rapidly rotating star, we need to ensure that the reference temperature that we use to set our models is defined in such a way that it is consistent and meaningful.  Instead of choosing a temperature corresponding to a specific location on the surface (such as the pole), we decide to define our reference temperature as the temperature that satisfies the Stefan-Boltzmann equation.  In this case, we calculate the luminosity and surface area terms via the triangulated mesh generated as part of the \spamms procedure.  In essence, this reference temperature represents an intensity weighted averaged surface temperature.

The choice of gravity darkening prescription can have an important effect on the distribution of the local temperature across the surface (see Fig. \ref{fig:surface}).  The two gravity darkening prescriptions most widely used in this mass regime are that of \citet{vonZeipel1924} and \citet{espinosalara2011}. While the Espinosa Lara prescription is considered to be more realistic, the von Zeipel prescription is still prevalent.  For this reason, we chose to use the Espinosa Lara prescription, however, we repeat the full analysis using the von Zeipel prescription in Appendix \ref{app:vz} for completeness.

\subsection{Comparison with spherical models}
Once the \spamms models accounting for the 3D deformations are computed, we compare them with \spamms models computed assuming spherical geometry.  In doing so, we ensure that any differences in the line profiles and the resulting observables are due solely to the change in geometry. In particular, we investigate whether the change in geometry has a significant effect on the observed temperature, observed helium abundance, and the observed rotation rates for each of the considered lines.  To do this, we calculate three values based on the resulting line profiles that are proxies for each of the physical parameters that we are investigating.

The temperature can be approximated based on the ratio of the equivalent widths (EW) of the helium lines, since in the temperature regime that we are considering, the ionization balance between \hea\ and \heb\ is very sensitive to the temperature.  Thus, by measuring the ratio of the EW of the \heb\ \l 4541 line to the \hea\ \l 4471 line, and comparing with non-rotating models, we can determine the observed temperature of each of the calculated \spamms models.  To calibrate this relation, we compute a small grid of non-rotating spherical \spamms models with temperatures ranging from 25kK to 50kK in steps of 1kK, assuming the same stellar parameters as outlined in Sec. \ref{subsec:spamms_setup}.  We then calculate the ratio of the EWs for each model in the non-rotating grid and use these values to interpolate the observed temperature for each of the rotating \spamms models.

The helium abundance can be approximated based on the sum of the EWs of the \hea\ \l 4471 and \heb\ \l 4541 lines.  A greater total EW corresponds to a higher helium abundance.  As with the temperature, we calibrate this relation by computing a small grid of non-rotating spherical \spamms models, but this time, we keep the temperature fixed at 40kK and we only allow the helium abundance to change.  We calculate 6 helium abundance steps: $N_\mathrm{He}/N_\mathrm{H} = 0.06, 0.10, 0.15, 0.2, 0.25, 0.30$.  We calculate the sum of the EWs for each model in the non-rotating grid and use these to interpolate the observed helium abundance for each of the rotating \spamms models.

Due to the temperature distribution and the resulting change in ionization balance across the surface of a rapidly rotating star, one may expect to measure different rotation rates for the \hea\ lines (which are formed primarily in the cooler equatorial region) and the \heb\ lines (which are formed primarily in the hotter polar region).  Differences in the measured rotation rates between these two lines can be approximated based on changes in the full width half maxima (FWHM) for each line.  Since the intrinsic non-rotationally broadened FWHM of the two lines are different, we do not compare the FWHM directly, but instead we compare the $\Delta$FWHM, which is the difference between the FWHM of the broadened line (FWHM$_{v\sin i}$) and the intrinsic FWHM of the non-rotating line (FWHM$_0$).  Thus, we calculate 

\begin{equation}
    \frac{\Delta\mathrm{FWHM}_\mathrm{He\textsc{ii}}}{\Delta\mathrm{FWHM}_\mathrm{He\textsc{i}}} = \frac{\mathrm{FWHM}_{v\sin i, \mathrm{He\textsc{ii}}} - \mathrm{FWHM}_{0, \mathrm{He\textsc{ii}}}}{\mathrm{FWHM}_{v\sin i, \mathrm{He\textsc{i}}} - \mathrm{FWHM}_{0, \mathrm{He\textsc{i}}}},
\end{equation}
as a proxy for the rotation rate.  By measuring the ratio of the $\Delta$FWHM of the \heb\ \l 4541 line to the \hea\ \l 4471 line, we can determine whether there is indeed a difference in the rotation rate between the two lines, and whether this effect is significant enough to be observed in nature. As with the previous two observables, we calibrate this using spherical \spamms models.  In this case, we keep the temperature fixed at 40kK and we set the helium abundance to $N_\mathrm{He}/N_\mathrm{H} = 0.2$.  We broaden the lines as is typically done when fitting spectra with 1D models, namely convolving the synthetic spectra with a rotational kernel matching the rotation rate.  From the imposed spherical assumption, both lines are forced to have the same rotation rate.

\section{Results}

\subsection{3D effects on measured temperature}

\begin{figure}[t]
    \centering
    \includegraphics[width=\linewidth]{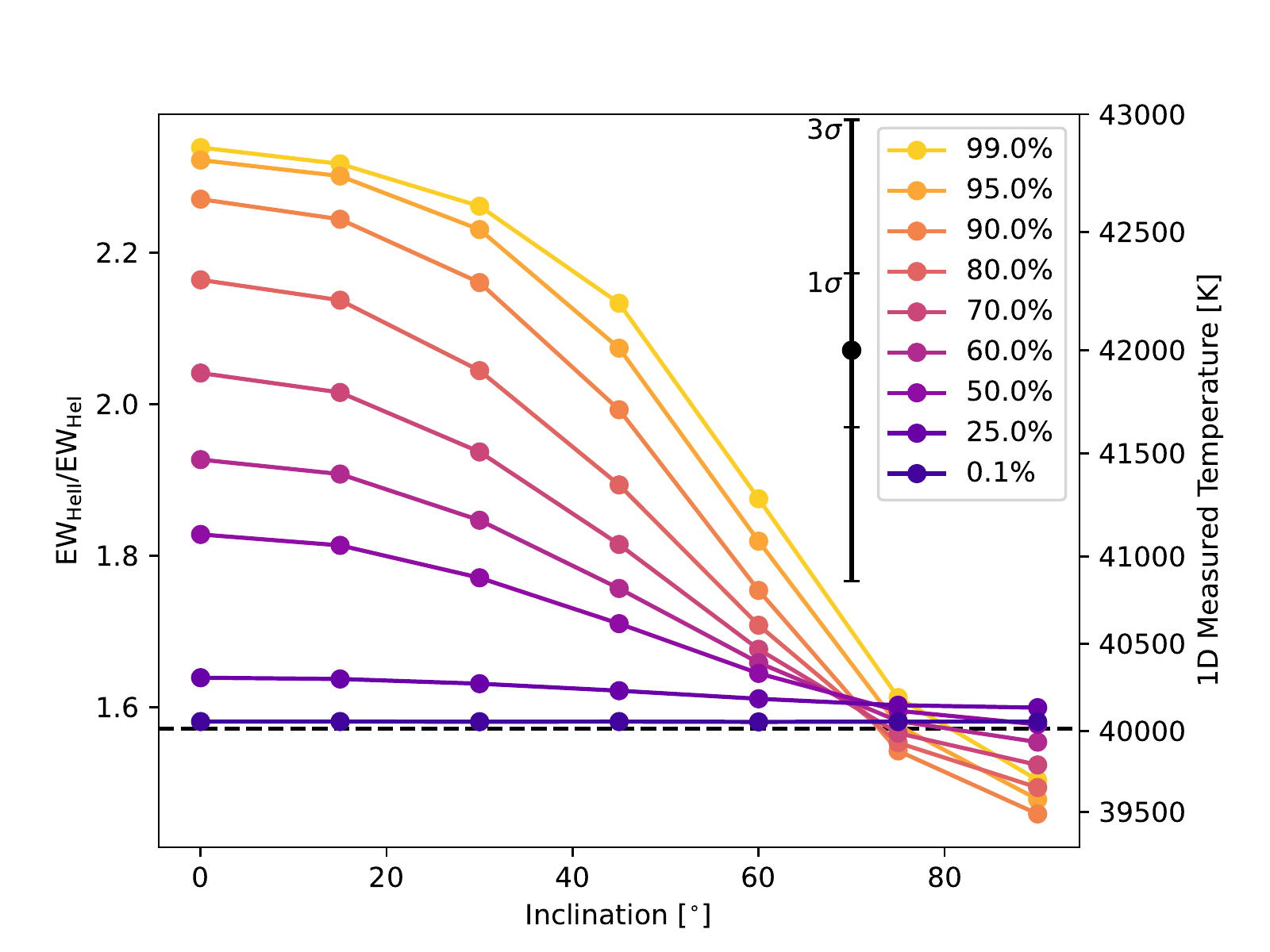}
    \caption{Ratio of the EWs of the \heb\ \l 4541 line to the \hea\ \l 4471 line as a function of inclination for a 40kK star at various inclinations, rotating at various rotation rates.  The rotation rate as a percentage of the critical rotation rate is indicated in the legend, with darker colors corresponding to lower rotation rates and lighter colors corresponding to higher rotation rates.  The EW ratio for a non-rotating model with the same stellar parameters is indicated with the horizontal dashed line. Simulated $1\sigma$ and $3\sigma$ error bars are indicated based on a simulated SNR of 300 and R of 6700.  The left y-axis indicates the ratio of the EWs while the right y-axis indicates the corresponding temperature.}
    \label{fig:teff}
    \end{figure}

Figure \ref{fig:teff} shows how the 1D measured temperature depends on the inclination and rotation rate.  We find that the ratio of the EWs of \heb\ to \hea\ (and thus the temperature) show a strong dependence on the inclination with higher inclinations corresponding to an underestimated EW ratio (lower 1D measured temperature) and lower inclinations corresponding to an overestimated EW ratio (higher temperature).  Furthermore, the degree of the deviation in the measured temperature is correlated with the rotation rate, with models closer to critical showing larger deviations than those with lower rotation rates.  In the most extreme cases, the difference between the measured temperature at an inclination of 0$^\circ$ and 90$^\circ$ spans almost 3.5kK ($\sim$10\% of the reference temperature).  

Putting this into perspective, when simulating a signal to noise ratio (SNR) of 300, a resolving power of $R = \lambda / \delta\lambda = $6700 \citep[R of the XSHOOTER UVB arm with 0.8" slit width;][]{Vernet2011} and performing a Monte Carlo simulation, the 1$\sigma$ standard deviation of the measured EW ratio results in a temperature error of $\sim$300K and a 3$\sigma$ error of $\sim$ 1kK.  If we instead assume a resolution of R = 40 000 \citep[R of the UVES Blue arm with 1.0" slit width;][]{Dekker2000} or R = 140 000 \citep[R of ESPRESSO HR mode;][]{Pepe2021}, the 3$\sigma$ errors become $\sim$ 0.5kK and $\sim$ 0.2kK respectively.  This implies that at a rotation rate approaching critical, the 1D measured temperature is more than 3$\sigma$ away from the true value for most inclinations.

\subsection{3D effects on measured helium abundance}

\begin{figure}[t]
    \centering
    \includegraphics[width=\linewidth]{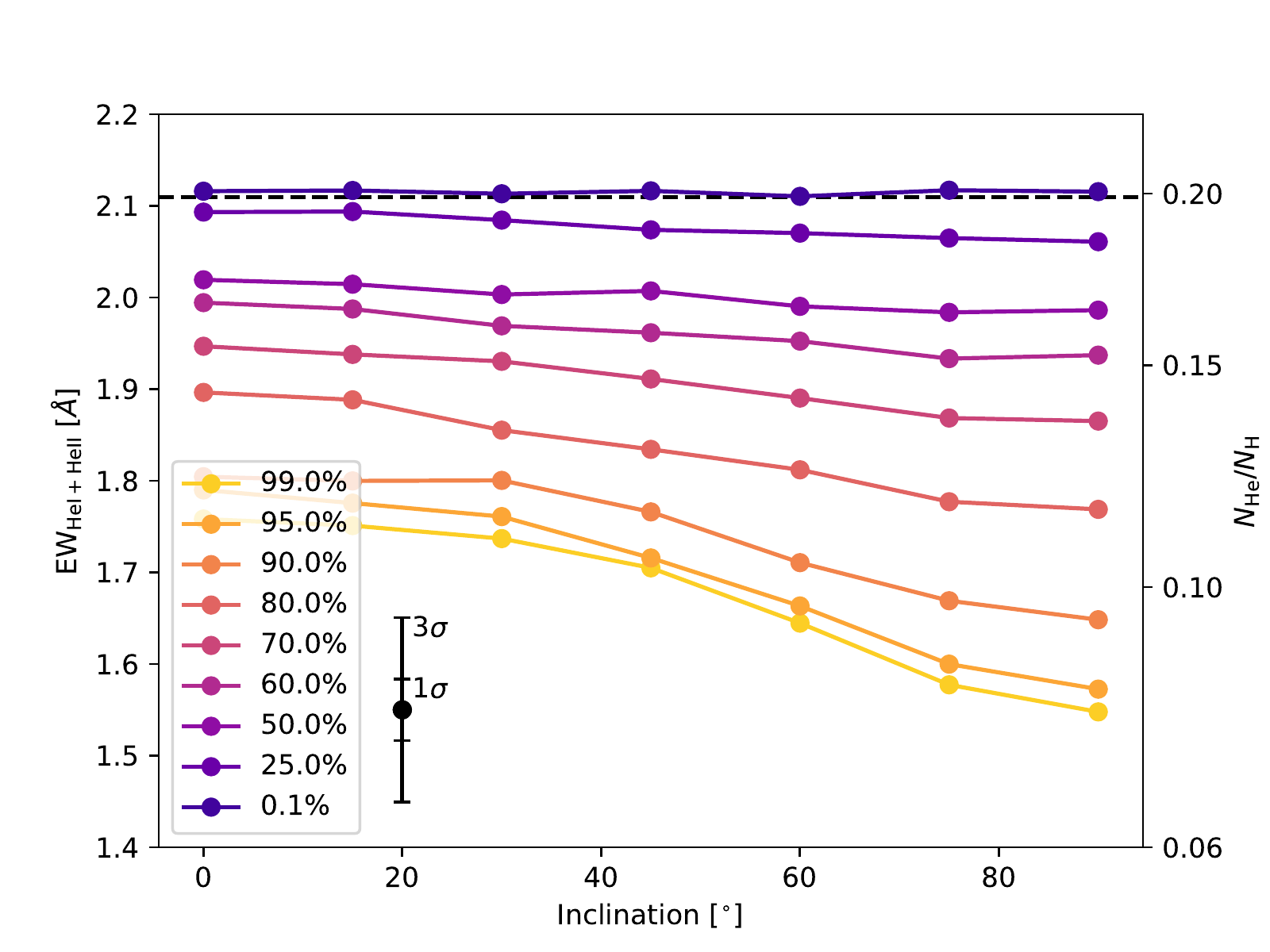}
    \caption{Same as Fig. \ref{fig:teff} but for the sum of the EWs of the \hea\ \l 4471 and \heb\ \l 4541 lines instead of the ratio the EW.  The left y-axis indicates the sum of the EWs while the right y-axis indicates the corresponding helium abundance.}
    \label{fig:abundance}
    \end{figure}

Figure \ref{fig:abundance} shows how the 1D measured helium abundance depends on the inclination and rotation rate for a star with a reference temperature of $T_\mathrm{ref}=$ 40kK.  In this case, the sum of the EWs shows a strong dependence on the rotation rate, resulting in an underestimation of helium abundance by $\sim$ 0.12 at a rotation rate of 99\% critical. Simulating a SNR of 300 and R of 6700, the expected error in the sum of EWs results in a 1$\sigma$ abundance error of $\sim$ 0.01 and a 3$\sigma$ error of $\sim$0.03. As with the temperature, increasing the resolving power will reduce the error bars even farther. This means that even at a relatively moderate rotation rate of 25\% critical, the 1D measured helium abundance can be more than 1$\sigma$ away from the true value and that at rotation rates of above 50\% critical, the measured value could be more than 3$\sigma$ away from the true helium abundance.  Unlike the temperature, however, the helium abundance shows a very moderate dependence on inclination, with the rotation rate being the dominant driver of the measured helium abundance.

    \begin{figure}[t]
    \centering
    \includegraphics[width=\linewidth]{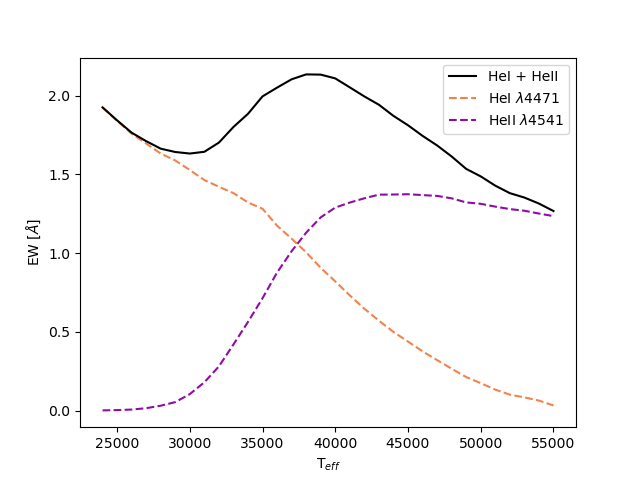}
    \caption{Sum of the EW of \hea\ \l 4471 and \heb\ \l 4541 as a function of temperature (solid black line). The individual contributions from each component are plotted in dashed lines as well (\hea\ is plotted in orange and \heb\ is plotted in purple)}
    \label{fig:EW}
    \end{figure}

This large discrepancy between the input and the 1D measured helium abundance is a consequence of the fact that the total EW of the helium lines is not constant as a function of temperature.   Figure \ref{fig:EW} demonstrates this effect. As the temperature increases and the ionization balance shifts from predominantly \hea\ to predominantly \heb, the total EW of the two considered lines changes significantly. A local maximum occurs at a temperature of $\sim$40kK, which corresponds with the intrinsic temperature we consider here.  Since the temperature varies across the surface of a rapidly rotating star, a majority of the surface will contribute \hea\ and \heb\ lines whose total equivalent widths are less than what is expected for a uniform spherical star with a temperature of 40kK.  Since spherical models assume that the conditions across the surface are constant, these models will drastically underestimate the helium abundance in this case as is shown in Fig. \ref{fig:abundance}.  Conversely, if one were to perform a similar experiment for a star with an intrinsic temperature of 30kK, corresponding to a local minimum in the relation, one would find that the helium abundances are systematically overestimated at all rotation rates when fitting with spherical models.

\subsection{3D effects on measured rotation rates}
\begin{figure}[t]
    \centering
    \includegraphics[width=\linewidth]{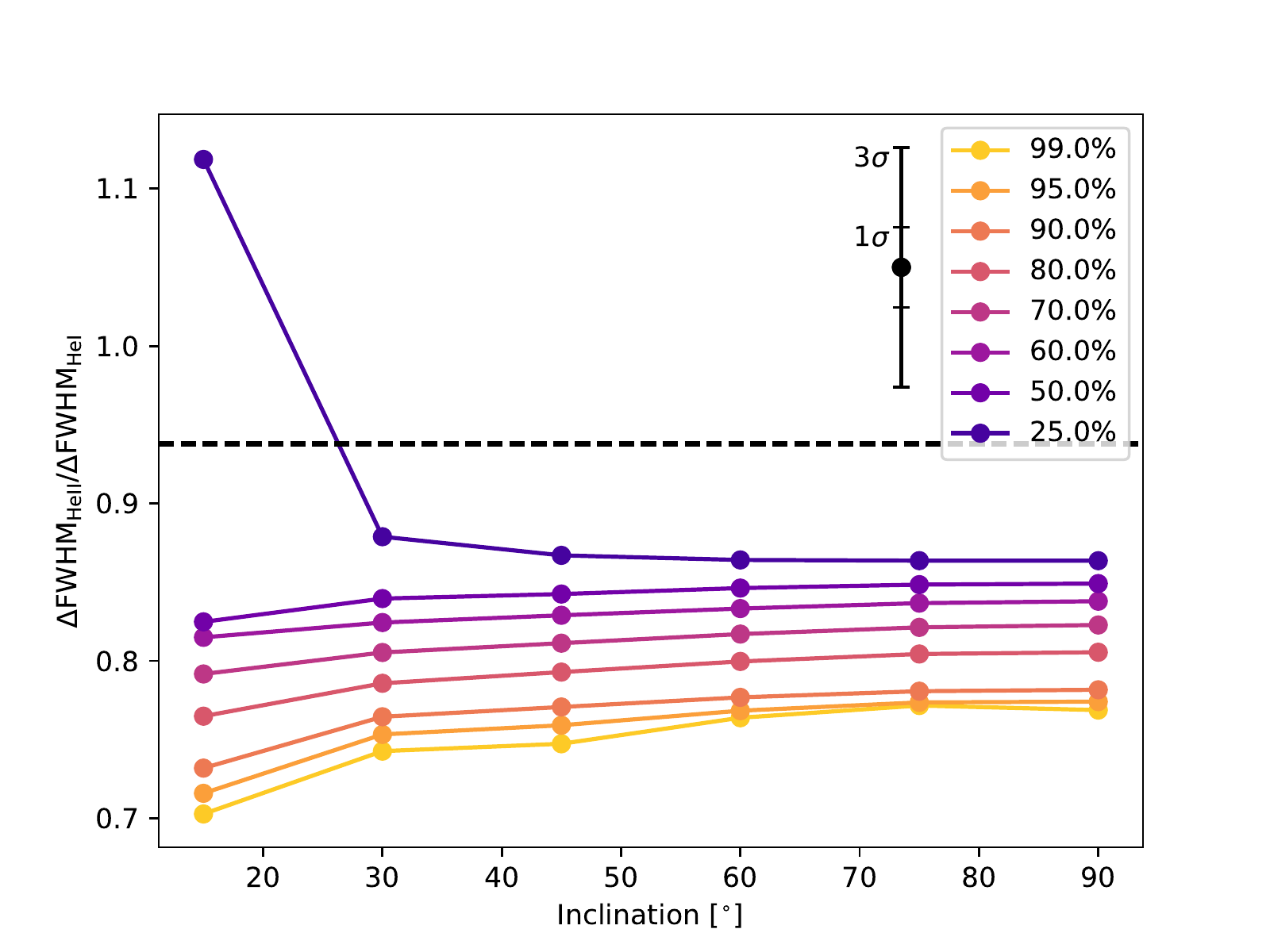}
    \caption{Same as Fig. \ref{fig:teff} but for the ratio of the $\Delta$FWHM of the \heb\ \l 4541 line to the \hea\ \l 4471 line instead of the ratio of the EW. The $\Delta$FWHM ratio for a spherical critically rotating model with the same stellar parameters is indicated with the horizontal dashed line.}
    \label{fig:FWHM}
    \end{figure}

Figure \ref{fig:FWHM} shows how the ratio of the $\Delta$FWHM of \heb\ to \hea\ varies with inclination and rotation rate.  Overall, the ratio of the $\Delta$FWHM shows a strong dependence on the rotation rate, with higher rotation rates showing a smaller ratio, and very little dependence on the inclination.  Due to the nature of the $\Delta$FWHM ratio, as the projected rotation rates approach 0, the ratio will approach infinity, so for this reason, inclinations of 0$^\circ$ are not included.  This is also the reason for the outlying point in the $\Upsilon$=0.25 model with an inclination of 15$^\circ$.  However, for the spherical case, our calibration shows that the ratio remains consistent to the third decimal for rotation rates between $\Upsilon$=0.1~to~1.0.  Both are plotted in Fig \ref{fig:FWHM}, however they appear indistinguishable on the scale of the plot, demonstrating that this ratio is both meaningful and constraining.

\section{Discussion}

\subsection{Inclination-independent measure of rotational velocity}
The determination of the rotation rate for a rapidly rotating single star is a notoriously difficult problem.  With 1D spectroscopic techniques, we are only able to measure the projected rotation rate, which is dependent on the inclination of the system.  To lift this degeneracy, in a binary system, one may assume that the inclination of the rapidly rotating star is aligned with the inclination of the binary (this can be a dangerous assumption as this is not always the case), however, for a single star system, no such inclination information is available.  Alternatively, if the system has a disk, as is the case for O/Be stars, inclination information can be obtained based on disk emission lines \citep{sigut2020}.

As we have demonstrated here, however, for a rapidly rotating massive star, there is a measurable difference between the rotation rate of the \hea\ and \heb\ lines.  Furthermore, the ratio of these rotation rates appears to provide a (nearly) inclination-independent measure of $\Upsilon$, breaking the degeneracy between the rotation rate and the inclination.  By measuring the rotation rates of different ionization states individually, this could allow for an independent measure of the rotation rate, and thus inclination, for systems that do not have a companion or a disk using spectroscopic data and 1D analysis techniques alone.  

\begin{figure}[t]
    \centering
    \includegraphics[width=\linewidth]{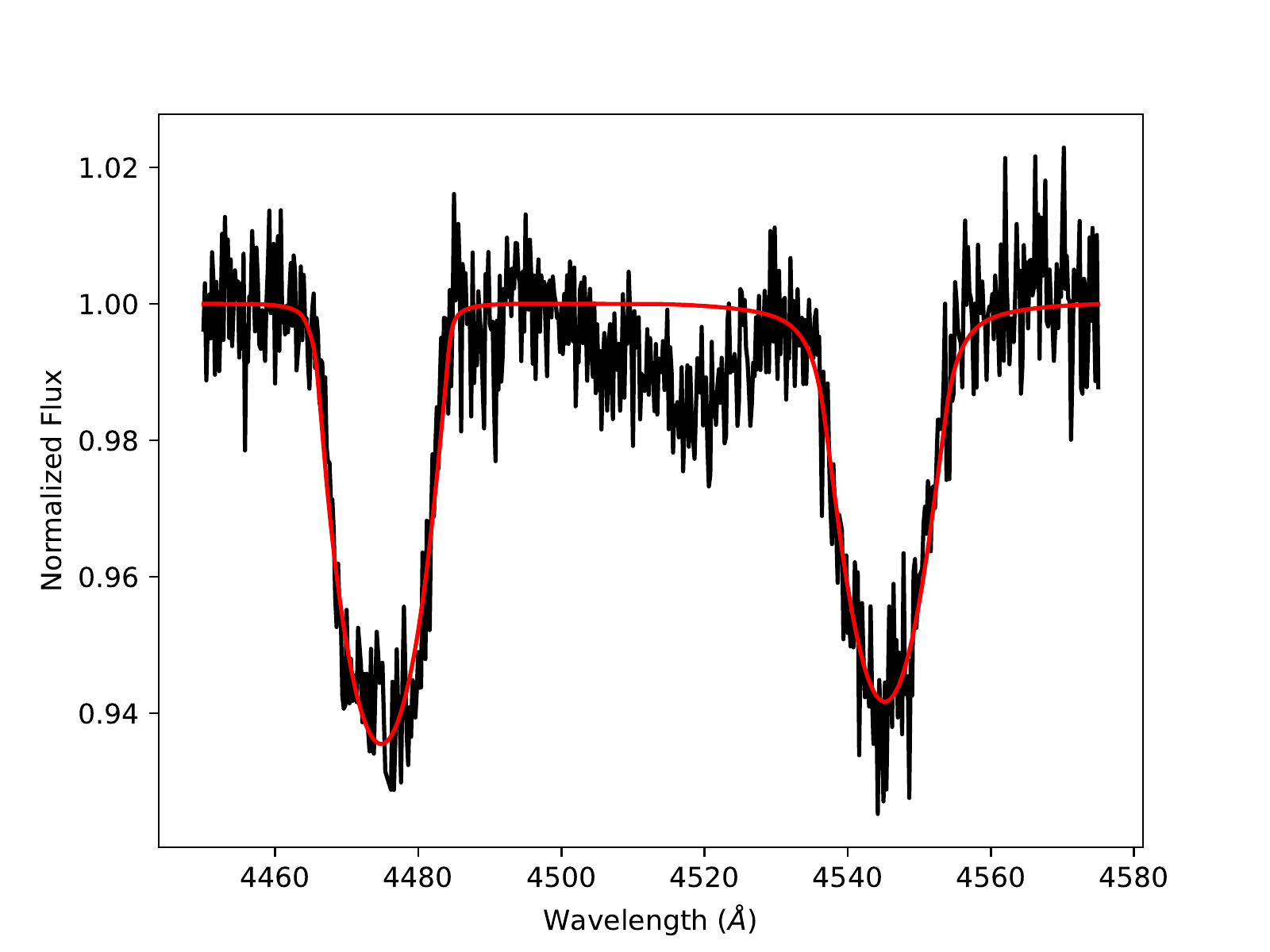}
    \caption{The observed spectrum of VFTS 285 is plotted in black and the best fit solution allowing for two different rotation rates for the \hea\ and \heb\ lines is plotted in red.}
    \label{fig:vfts285}
    \end{figure}

As a demonstration that this effect is observable, we perform a simple fit on XSHOOTER data of the known rapidly rotating O-type system VFTS 285 \citep[$v\sin i = 609 \pm 29$ \kms; ][]{Ramirez-Agudelo2013, Shepard2020, Shepard2022} using a grid similar to the ones generated in Sect. 2.2. Allowing for a different rotation rate between the \hea\ and \heb\ lines and optimizing over temperature, helium abundance and the projected rotation rates, we find a temperature of $T_\mathrm{eff} = 32.6 \pm 1.9$ kK, helium abundance of $N_\mathrm{He}/N_\mathrm{H} = 0.14 \pm 0.06$ and rotation rates for \hea\ of $v_\mathrm{HeI}\sin i = 613 \pm 16$ \kms and for \heb\ of $v_\mathrm{HeII}\sin i = 567 \pm 32$ \kms (see Fig. \ref{fig:vfts285}).  The rotation rates are not consistent within error with one another and differ by $\sim$50~\kms, and as expected, the \hea\ lines show a higher rotation rate than \heb.  In order to convert this difference in measured rotation rates to $\Upsilon$, a full calibration of the parameter space (including helium abundance, temperature, rotation rate and inclination) is required, however, this is beyond the scope of the current work.  We note that the SNR of the observed data is only $\sim$140 and we are still able to measure this difference indicating that the error bars in Sect. 3.3 may be overly conservative.  This likely results from the use of FWHM as opposed to rotation rate.

In practice, rotation is not the only cause of broadening in massive stars, so other broadening mechanisms may complicate these measurements, however techniques have been developed to distinguish between the different broadening mechanisms \citep[see e.g., ][]{Simon-Diaz2014, Simon-Diaz2017}. Expanding one's analysis to include other elements with multiple ionization states may further aid in overcoming this issue.

\subsection{Confronting rotational mixing}
As discussed in Sect. \ref{sec:intro}, rotation is expected to lead to internal mixing via sheer instabilities and Eddington-Sweet circulations.  A consequence of this internal mixing is that products of nuclear burning are expected to be brought to the surface.  In the core of a massive star, hydrogen is fused into helium via the CNO cycle, leading to an increase in the Helium abundance.  While the total number of carbon, nitrogen and oxygen atoms are expected to remain the same, differences in the reaction rates of the different steps in the CNO cycle cause the abundances of these three elements to change.  The nitrogen abundance is expected to increase, while the carbon and oxygen abundances are expected to decrease.  The abundance changes of He, C, N and O are expected to be visible on the surface of the star as it mixes \citep{Maeder2000a, Maeder2000b, Brott2011a}. Furthermore, evolutionary models suggest that the effective temperature is also expected to increase \citep{Yoon2005, Yoon2006}.  For these reasons, effective temperature and abundance anomalies are typically used to constrain the internal mixing properties of massive stars \citep[e.g., ][]{Hunter2008}.

As we have demonstrated here, however, both of these diagnostics are very heavily influenced by the 3D geometry of the system.  For the most rapidly rotating systems, the measured temperature can vary by up to 3.5kK depending on the inclination, and can be up to 3kK away from the reference temperature.  Depending on the age of the system, these variations can be on the same scale as what is expected from internal mixing.  Furthermore, the measured abundances can also be significantly underestimated or overestimated depending on the temperature distribution across the surface, and the sum of EWs curve for the chosen spectral lines (see Fig. \ref{fig:EW}). 

These effects are currently not well accounted for in the diagnostic plots that we currently use to study internal mixing.  The Hunter plot \citep[see Fig. 1 in ][]{Hunter2008}, which shows the observed nitrogen abundance as a function of projected rotation rate has become a popular way to study the rotational mixing.  Aside from failing to distinguish between low inclination rapidly rotating systems and high inclination moderately rotating systems, the 3D effects on the nitrogen abundance are not treated.  Accounting for the 3D geometry would allow us to mitigate both of these issues, and may help explain some of the anomalous regions in the Hunter plot.  Another important example is the Hertzsprung-Russell (HR) diagram.  Given the dependence of 1D measured temperature on the inclination, the placement of a rapidly rotating star on the HR diagram can lead to an incorrect interpretation of its evolutionary status.  Unfortunately, the spectroscopic HR diagram and the color-magnitude diagram suffer from the same issues.  Accounting for the 3D geometry should allow for a more accurate placement of rapidly rotating stars within these diagrams.

\section{Conclusions}

We have presented an investigation of the importance of 3D geometry in the determination of spectroscopic parameters for O-type rapidly rotating stars.  Using synthetic spectra accounting for the 3D geometry generated using \textsc{spamms}, we have shown that the measured temperature and surface abundances can be more than 3$\sigma$ away from the input values when using 1D analysis techniques.  The measured temperature shows a strong dependence on both the inclination and the rotation rate, while the He surface abundance appears to depend more critically on the rotation rate. In the most extreme case, the measured temperature can vary by up to 10\% depending on the inclination of the system, and the He abundance can be either overestimated or underestimated by more than a factor of 2.  We have also demonstrated that the difference in rotation rate between different ionization states of the same element may be used to constrain the rotation rate, independent of the system's inclination. 

Overall, we have shown that the 3D geometry of rapidly rotating stars can have a significant effect on the spectra and therefore cannot be neglected.  Without accounting for the 3D geometry of rapidly rotating systems, we may be unintentionally biasing our observational understanding of their evolutionary state and miscalibrating our theoretical models of internal mixing.

\begin{acknowledgements}
  I thank A. Escorza, J. Bodensteiner, D.M. Bowman and S. Simon-Diaz for their useful discussion on this topic.  In addition to the codes explicitly cited in the text, this paper makes use of the following packages: \textsc{matplotlib} \citep{Hunter2007}, \textsc{numpy} \citep{Harris2020}, \textsc{scipy} \citep{Virtanen2020}, \textsc{astropy} \citep{AstropyCollaboration2013, AstropyCollaboration2018, AstropyCollaboration2022}, and \textsc{schwimmbad} \citep{Price-Whelan2017}.
\end{acknowledgements}

\bibliographystyle{aa}
\bibliography{mybib}

\begin{appendix}

\section{von Zeipel gravity darkening case}\label{app:vz}

\begin{figure*}[t]
    \centering
    \includegraphics[width=\linewidth]{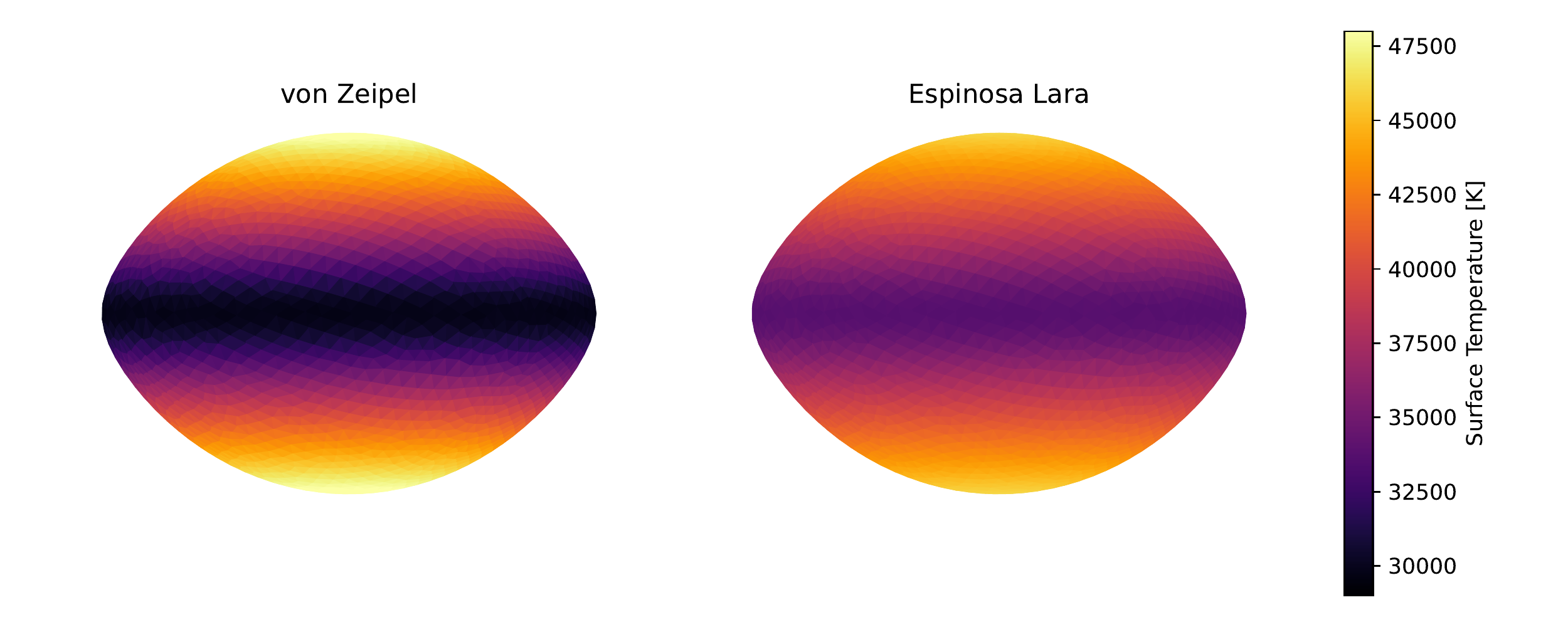}
    \caption{Comparison of the surface temperature structure of a 40kK star rotating at 90\% critical assuming von Zeipel gravity darkening (left) and Espinosa Lara gravity darkening (right).}
    \label{fig:surface}
    \end{figure*}

As stated in the main text, the choice of gravity darkening prescription can have a large impact on the distribution of the temperature across the surface of a deformed star.  Here we consider the von Zeipel (vZ henceforth) gravity darkening prescription.  In contrast with the Espinosa Lara (EL henceforth) prescription, the vZ prescription shows a much stronger dependence between the temperature and surface gravity.  Thus in highly deformed stars, as the surface gravity approaches zero, so does the surface temperature.  This leads to a wider distribution of temperatures across the surface when compared to the EL prescription.  Figure \ref{fig:surface} demonstrates how the two prescriptions compare for a 40 kK star rotating at 90\% critical.  That said, at lower rotation rates, the divergence between these two prescriptions shrinks.


\begin{figure}[t]
    \centering
    \includegraphics[width=\linewidth]{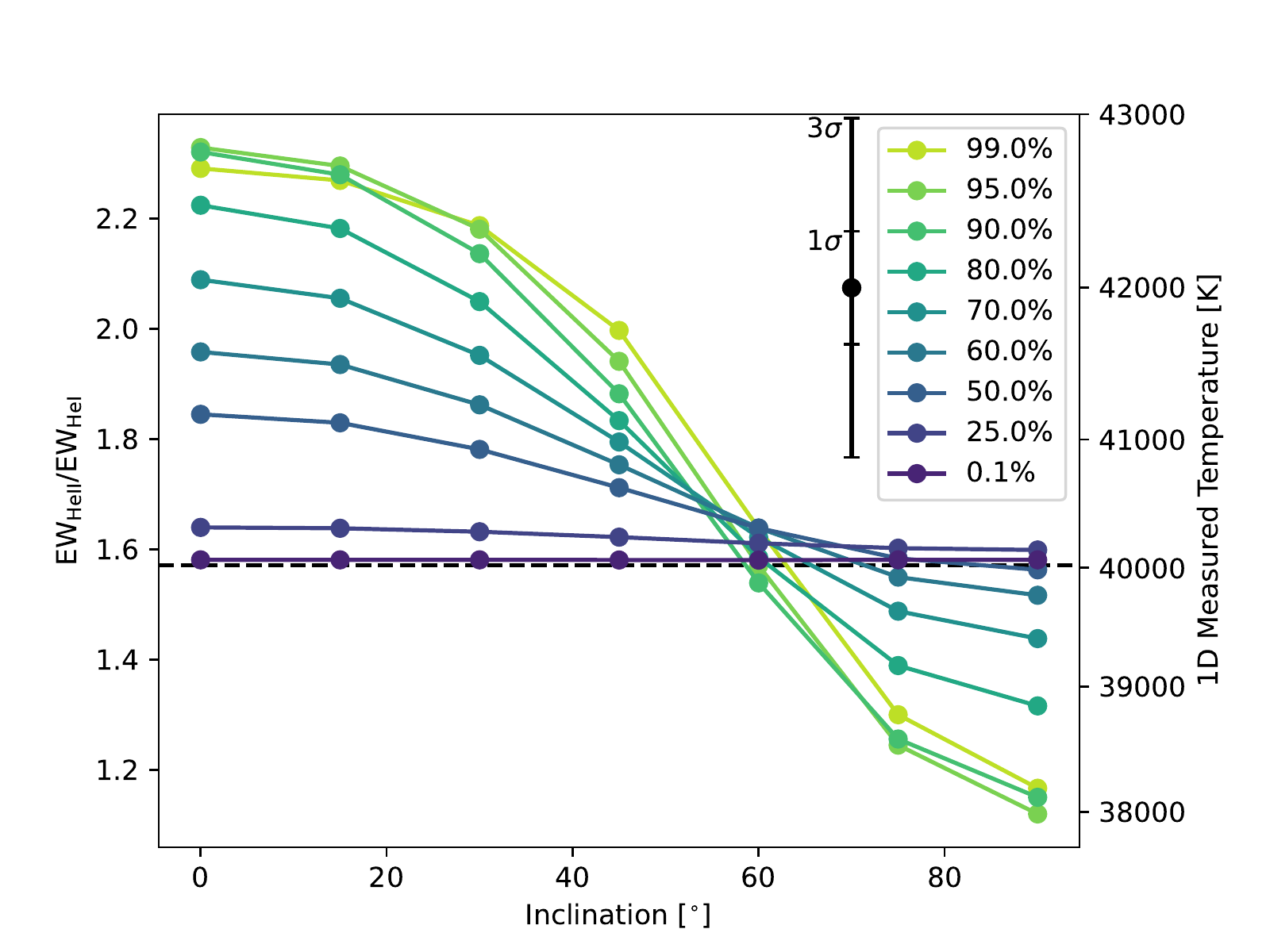}
    \caption{Same as Fig. \ref{fig:teff}, but assuming the von Zeipel gravity darkening prescription instead of the Espinosa Lara}
    \label{fig:teff_vZ}
    \end{figure}

\begin{figure}[t]
    \centering
    \includegraphics[width=\linewidth]{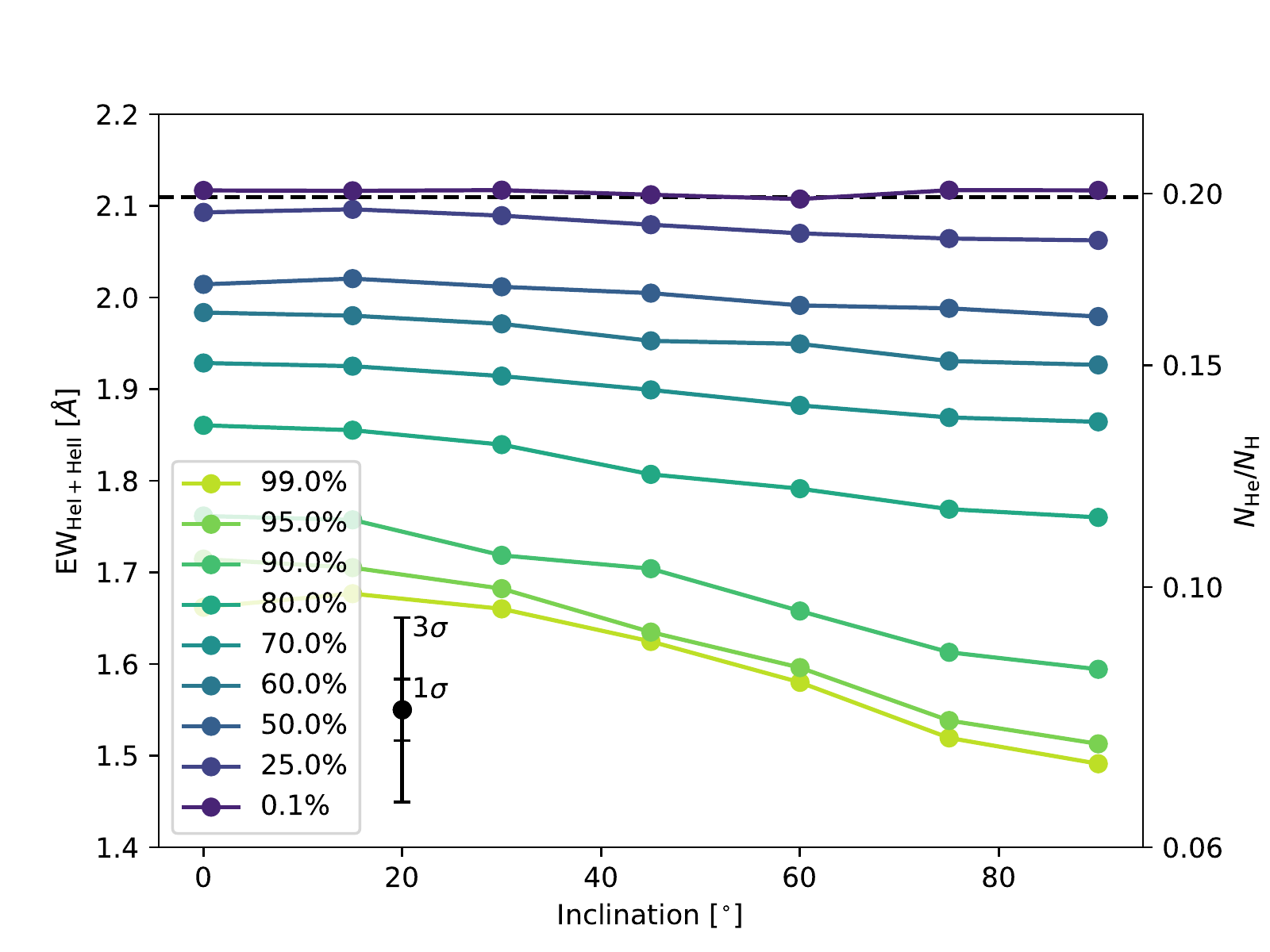}
    \caption{Same as Fig. \ref{fig:abundance}, but assuming the von Zeipel gravity darkening prescription instead of the Espinosa Lara}
    \label{fig:abundance_vZ}
    \end{figure}

\begin{figure}[t]
    \centering
    \includegraphics[width=\linewidth]{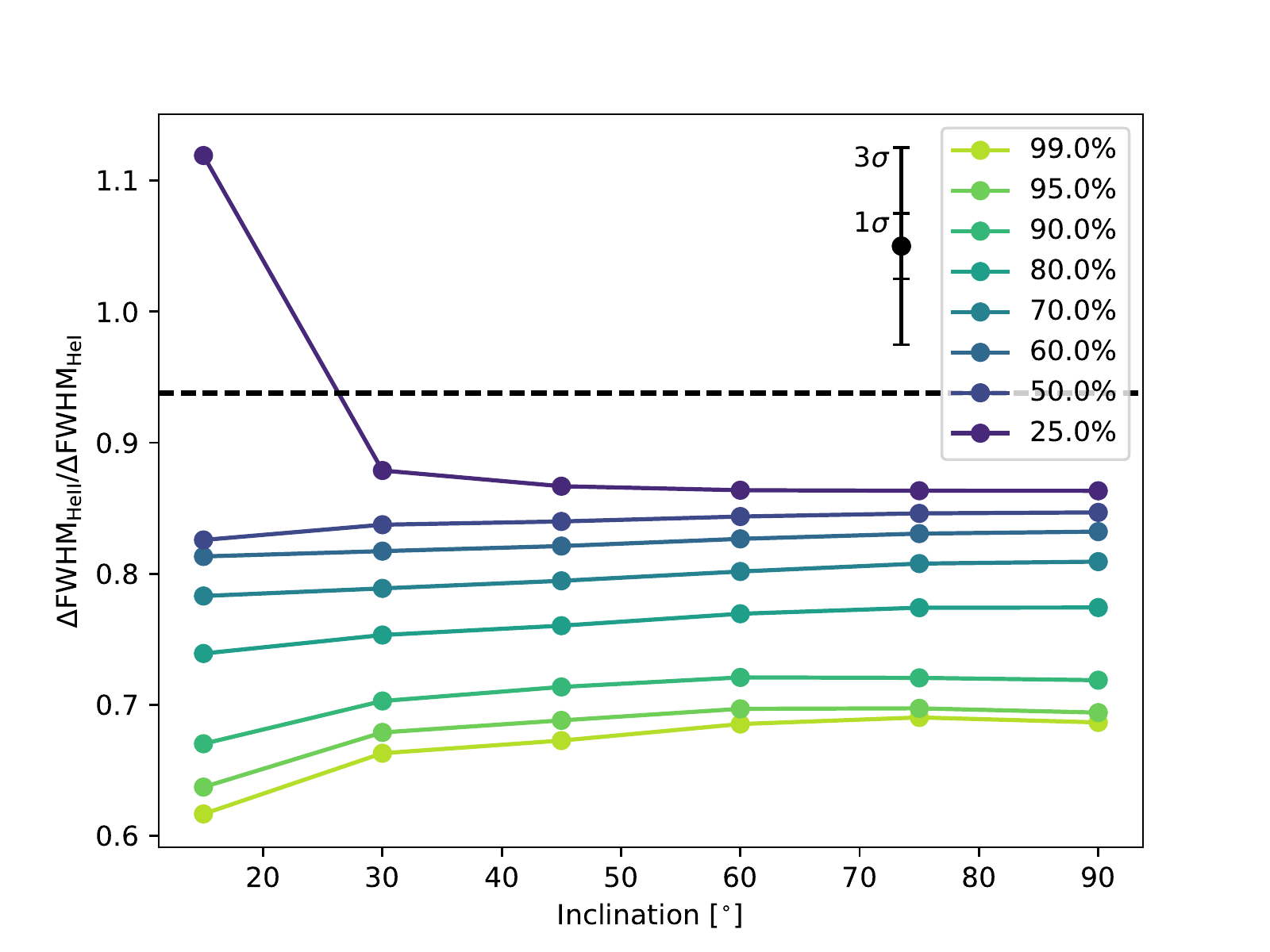}
    \caption{Same as Fig. \ref{fig:FWHM}, but assuming the von Zeipel gravity darkening prescription instead of the Espinosa Lara}
    \label{fig:FWHM_vZ}
    \end{figure}

Here, we repeat the same analysis outlined in the main text, however we do so using the vZ gravity darkening prescription instead.  Figures \ref{fig:teff_vZ}, \ref{fig:abundance_vZ} and \ref{fig:FWHM_vZ} are vZ equivalent plots to Figs. \ref{fig:teff}, \ref{fig:abundance} and \ref{fig:FWHM} in the main text.

Qualitatively the vZ results are very similar to the EL results, indicating that our conclusions are indeed sound and independent of the chosen gravity darkening prescription.  The magnitude of the effects vary slightly between the different plots, but the overall picture is very consistent.  That said, the biggest difference in results between the two prescriptions can be seen in the temperature plots (Figs. \ref{fig:teff} and \ref{fig:teff_vZ}).  We can see that the two prescriptions diverge at higher inclinations, while they remain almost indistinguishable at lower inclinations. In the 99\% critical case we can see that the difference between the 1D measured temperature at an inclination of 0$^\circ$ and 90$^\circ$ is closer to 5kK for vZ as opposed to 3.5 kK for EL.  Interestingly, the inclination where the 1D measured temperature matches the reference temperature changes between the two gravity darkening prescriptions: $\sim70^\circ$ for EL and $\sim60^\circ$ for vZ. 

For the other two plots, the shapes of the curves are almost indistinguishable between the two prescriptions.  The effect on the measured helium abundance is marginally more pronounced for vZ (helium abundance is underestimated by 0.13 for vZ instead of 0.12 for EL for the 99\% critical case), but the trend with inclination is in very good agreement.  Similarly, for the ratio of the $\Delta$FWHM, the vZ prescription is more pronounced, with a larger dispersion between the 25\% and 99\% critical cases, but aside from that, the two prescriptions are virtually the same.

\end{appendix}
\end{document}